\documentstyle[11pt]{article}
\newcommand{\rf}[1]{(\ref{#1})}
\newcommand{\bea}{\begin{eqnarray}}
\newcommand{\eea}{\end{eqnarray}}
\newcommand{\e}{\rm e}

\renewcommand{\l}{\lambda}

\newcommand{\del}{\delta}
\newcommand{\ep}{\varepsilon}
\newcommand{\om}{\omega}
\newcommand{\sg}{\sigma}

\newcommand{\oh}{\frac{1}{2}}
\newcommand{\oq}{\frac{1}{4}}

\newcommand{\tr}{{\rm Tr}\,}
\newcommand{\ra}{\right\rangle}
\newcommand{\la}{\left\langle}

\newcommand{\hK}{\hat{\cal K}}
\newcommand{\noi}{\vspace{12pt}\noindent}
\newcommand{\no}{\nonumber}
\newcommand{\non}{\no\\}
\def\void{}
\def\labelmark{}

\newenvironment{formula}[1]{\def\labelname{#1}
\ifx\void\labelname\def\junk{\begin{displaymath}}
\else\def\junk{\begin{equation}\label{\labelname}}\fi\junk}%
{\ifx\void\labelname\def\junk{\end{displaymath}}
\else\def\junk{\end{equation}}\fi\junk\labelmark\def\labelname{}}

{\ifx\void\labelname\def\junk{\end{array}\end{displaymath}}
\else\def\junk{\end{array}\right.\end{equation}}
\fi\junk\labelmark\def\labelname{}\def\junk{}
}

\newcommand{\beq}{\begin{formula}}
\newcommand{\eeq}{\end{formula}}
\newcommand{\beqv}{\begin{formula}{}}

\begin{document}
\topmargin 0pt
\oddsidemargin 5mm
\headheight 0pt
\headsep 0pt
\topskip 9mm

\hfill    NBI-HE-96-xx

\hfill June  1996

\begin{center}
\vspace{24pt}
{\Large \bf New universal spectral correlators}

\vspace{24pt}

{\sl J. Ambj\o rn }  and {\sl G. Akemann}\footnote{Address 
after June 96: Inst.f.Theor.Physik, Appelstr.2, D-30167 Hannover, Germany}

\vspace{6pt}

 The Niels Bohr Institute\\
Blegdamsvej 17, DK-2100 Copenhagen \O , Denmark\\

\vfill
\end{center}

\begin{center}
{\bf Abstract}
\end{center}

\vspace{12pt}

\noindent
We study the universal properties of distributions of eigenvalues
of random matrices in the large $N$ limit. The distributions fall 
in universality classes characterized entirely by the support of
the spectral density.
\vfill

\newpage

\section{The theorems}

Random matrices have found a wide range of applications in
solid state physics, nuclear physics and high energy physics.
An important observable in most of these applications
is the correlator between eigenvalues of the random matrices.
In addition to being random, the matrices are often large, and
it is natural to take the so called large $N$ limit
in many of the applications. The general set up is as follows:
we consider an ensemble of random matrices
\beq{*1}
P(\phi) = \frac{1}{Z} {\e}^{-N \tr V(\phi)}, ~~~~
Z= \int d \phi \; {\e}^{-N \tr V(\phi)},~~~
V(\phi)=\sum_{k} \frac{g_k}{k} \phi^k
\eeq
and in this ensemble we can ask for expectation values of a
certain observables :
\beq{*2}
\la f(\phi) \ra = \int d\phi \; P(\phi) \; f(\phi).
\eeq
In the following we will assume that the random matrices
$\phi$ are $N \times N$ Hermitian matrices, but similar
results can be proven for ensembles of complex matrices matrices.

\noi
The one- and two-point ``resolvents''  are defined by
\bea
G(z) &=& \frac{1}{N} \,\la  \tr \frac{1}{z - \phi} \ra \label{*3}\\
G(z_1,z_2) &=& \la \tr \frac{1}{z_1 - \phi}
\;\tr \frac{1}{z_2 - \phi} \ra -
\la \tr \frac{1}{z_1 - \phi} \ra\la \tr \frac{1}{z_2 - \phi} \ra,\label{*4}
\eea
They are related to the spectral density and the correlator of densities
\beq{*5}
\rho(\l) \equiv \la S(\l) \ra,~~~~~~S(\l) \equiv
\frac{1}{N} \, \sum_{i=1}^N\del (\l -\l_i)
\eeq
and
\beq{*6}
\rho (\l,\l') \equiv
\la S(\l) S(\l')\ra - \la S(\l)\ra\la S(\l')\ra
\eeq
in the following way
\beq{*7}
\rho(\l) = \frac{-1}{2\pi i} \,\Bigl( G(\l + i\ep ) -G(\l-i \ep)\Bigr)
\eeq
and
\beq{*8}
\rho(\l,\l') = \frac{1}{N^2}\Bigl(\frac{-1}{2 \pi i}\Bigr)^2 \Bigl(
G(+,+) + G(-,-) -G(+,-)-G(-,+) \Bigr),
\eeq
where $G(\pm,\pm)\equiv G(\l\pm i\ep,\l'\pm i\ep')$.

\noi
The following theorem was proven in \cite{ajm}
(and partly rediscovered by the authors of \cite{bz},
after whom it is sometimes called Brezin--Zee universality):

\begin{itemize}
\item[~]
{\bf Theorem 1:} Assume that the spectral density $\rho(\l)$
has support in a single interval in the large $N$ limit.
Then the two-point correlator (and thus $\rho(\l,\l')$) is
{\it universal}, i.e. independent of $V(\phi)$, and given by
(in the case of a symmetric potential, such that the
support is $[-a,a]$)
\bea
G(z_1,z_2) &=& \frac{1}{4(z_1-z_2)^2}\left(-2+
\frac{(z_1^2-a^2)+(z_2^2-a^2)}{\sqrt{(z_1^2-a^2)(z_2^2-a^2)}}\right) 
\non
&& -\oq \frac{1}{\sqrt{(z_1^2-a^2)(z_2^2-a^2)}}.
\label{*9}\eea
\end{itemize}

\noindent
For the general formula for an arbitrary potential, as well as
for complex matrices, rather than Hermitian matrices,
we refer to \cite{ajm}.
In fact much more was proven in \cite{ajm}. Any multi--point resolvent
\beq{*10}
G(z_1,\ldots,z_n) \equiv N^{2n-2}\la \frac{1}{N} \tr \frac{1}{z_1-\phi} \cdots
\frac{1}{N}\tr \frac{1}{z_n-\phi} \ra_{conn},
\eeq
where {\it conn} refers to the connected part of the multi--point
resolvent, is also universal. Closed and very simple expressions were
given for the multi--point resolvents in the case where matrix
ensemble consisted of complex matrices. Note that the generalization
of \rf{*8} is
\beq{*8a}
\la S(\l_1) \cdots S(\l_n) \ra_{conn} =
\frac{1}{N^{2n-2}} \Bigl(\frac{-1}{2\pi i}\Bigr)^n
\sum_{\sg_k=\pm 1} (-1)^{\Sigma_k \sg_k} G(z_1+\sg_1 i \ep,\ldots,
z_n + \sg_n i \ep).
\eeq

\noi
In addition it is possible to develop a systematic $1/N^2$ expansion:
\bea
G(z_1,\ldots,z_n) &=& \sum_{h=0}^\infty \frac{1}{N^{2h} }
\; G_h(z_1,\ldots,z_n), \label{*add1}\\
F &=&   \sum_{h=0}^\infty \frac{1}{N^{2h} }
\;F_h.  \label{*add2}
\eea
Again the two--point
resolvents and in this case even the corrections to the spectral density
$\rho(\l)$ as well as the corrections to the free energy $N^2 F = \log Z$
are all universal \cite{ackm}. These considerations have
later been extended to supermatrices \cite{superm}.

\noi
In the above discussion universality means that the resolvents can
be presented in a {\it form} independent of the potential. Of course
the function \rf{*9} depends on the potential through
the endpoints  of the eigenvalue distribution $\rho(\l)$.
The same remark is true for the multi--point resolvents and for higher
order expansions in $1/N^2$. However, for the $n$--point resolvent
calculated to non-trivial order $h$ in the $1/N^2$ expansion only 
$2(3h-2 +n)$ additional parameters $M_{1,2}^{(l)}$ are involved for 
an arbitrary potential (see def. \rf{*23}).
Yet, some of these might vanish, as will be the case
for a Gaussian potential where  all additional parameters vanish but one.
The somewhat misleading statement found in \cite{bz} that
all multi--point density correlators vanish to order $1/N^n$ is a trivial
fact which follows from the factorization property of
the large $N$ expansion.
The first non-trivial term is of order $1/N^{2n-2}$
for the $n$-point density correlator \rf{*8a}
and it is explicitly non-Gaussian in
nature for $n >2$ because of the additional parameters. 
These observations can be summarized as follows
\cite{ackm}:

\begin{itemize}

\item[~]
{\bf Theorem 2:} If the spectral density in the large $N$ limit
has support $[-a,a]$ all higher $1/N^2$ corrections
to the free energy $F$, to the $n$-point resolvents, to the
spectral density and any multiple eigenvalue correlators are universal.
The correction to the $n$-points resolvents are of the form
\beq{*G}
G_h(z_1,\ldots,z_n)\ = \
\frac{R_h\Big( \{z_i^2-a^2\},\{M_a^{(l)}\},a\Big)}
{ \sqrt{ \prod_{i=1}^{n} (z^2_i-a^2)}},
\eeq
where the $R_h$ are simple rational functions of the arguments,
which can be determined iteratively in the  $1/N^2$ expansion.
\end{itemize}

\noi
We refer to \cite{ackm} for details, generalizations to the
situation where the support is not symmetric and to
\cite{akm} for complex matrix models.

\noi
All these considerations have been based on the assumption that
$\rho(\l)$ has support in one  interval on the real axis.
More general situations can occur for complicated potentials,
and in fact it is natural for applications in solid state
physics to have such a situation since the eigenvalues of $\phi$
are viewed as (part of) the  eigenvalues of a Hamiltonian which
can have a band structure. In this paper we prove

\begin{itemize}
\item[~]
{\bf Theorem 3:} Assume that the support of the spectral density $\rho(\l)$
in the large $N$ limit consists of $s$ intervals $[x_{2j},x_{2j-1}]$,
$j=1,\ldots,s$. The large $N$ limit of the two-point
resolvents $G_0(z_1,z_2)$
(and therefore the correlators of eigenvalues)
fall in universality classes characterized by $s$ and they depend
only on the potential $V$ through the endpoints $x_i$.
For each $s>1$ it is possible to generalize the statements
in Theorem 1 and 2 concerning universality.
\end{itemize}

\noi
Let us here present the explicit solution, analogous to \rf{*9},
in the case where the support of  $\rho(\l)$ consists of two
intervals $[x_4,x_3]$ and $[x_2,x_1]$, $x_4 < x_3 < x_2 < x_1$,
and where we for simplicity assume that $V(\phi)$ is an even
function of $\phi$ such that $x_4 = -x_1$ and $x_3=-x_2$.
If we denote $x_1,x_2$ by $a,b$ we have
\bea
G_0(z_1,z_2) &=& \frac{1}{4(z_1-z_2)^2}
\left(-2+\frac{(z_1^2-a^2)(z_2^2-b^2)+(z_1^2-b^2)(z_2^2-a^2)}{
           \sqrt{(z_1^2-a^2)(z_1^2-b^2)(z_2^2-a^2)(z_2^2-b^2)}}\right) \non
&& + \frac{(a+b)^2}{4} \frac{E(k)}{K(k)}
\frac{1}{\sqrt{(z_1^2-a^2)(z_1^2-b^2)(z_2^2-a^2)(z_2^2-b^2)}} ,
\label{*11}
\eea
where $k = 2\sqrt{ab}/(a+b)$,
and where $E(k)$ and $K(k)$ denote the complete elliptic
integrals of first and second kind.

\noi
In the next section we will briefly outline how to prove theorem 3.

\section{Proofs}

The basic tool for proving the above statements is the loop
equation in the following form (\cite{ackm}):
\beq{*12}
\hK G (z) = G^2(z) + \frac{1}{N^2} G(z,z).
\eeq
In this formula  $\hK$ is a linear operator
\beq{*13}
(\hK f)(z) \equiv \oint_C \frac{d\om}{2\pi i} \frac{V'(\om)}{z-\om} f(\om),
\eeq
and the contour encloses all singularities of $G(z)$, but not $z$.
Note that the support $\sg(\l)$ of
$\rho(\l)$ coincides with the singularities
of $G(z)$. In large $N$ limit they will be located at $s$ cuts on the real
axis. If the $x_i$, $i=1,\ldots,2s$ denote the positions of the endpoints
of the  cuts ($x_1 > x_2 > \cdots$) the solution to \rf{*12}
in the large $N$ limit can be
written in closed form since we can discard the term
involving $G(z,z)$. We get
\beq{*15}
G_0 (z) = \oh \oint_C \frac{d\om}{2\pi i} \frac{V'(\om)}{z-\om}
\sqrt{\prod_{i=1}^{2s} \frac{z-x_i}{\om-x_i}}.
\eeq
In addition the endpoints of the cuts are uniquely determined
by the fact that $G_0(z) \sim 1/z$ for large $|z|$, and a certain
stability requirement \cite{david}. The first requirement leads to
$s+1$ equations:
\beq{*16}
\oh \oint_C \frac{d\om}{2\pi i} 
\frac{V'(\om) \om^k}{\sqrt{\prod_{i=1}^{2s} (\om-x_i)}} 
= \del_{k,s},~~~~~k=0,\ldots,s.
\eeq
If we define the polynomial $M(z)$ by
\beq{*17}
M(z) =
\oint_{C_\infty} \frac{d\om}{2\pi i} \frac{V'(\om)}{(\om-z)
\sqrt{\prod_{i=1}^{2s} (\om-x_i)}},
\eeq
where $C_\infty$ is a contour at infinity in the complex plane, the
requirement of stability implies \cite{jurek}
\beq{*18}
\int^{x_{2k}}_{x_{2k+1}} d\l \;M(\l)
\sqrt{\prod_{i=1}^{2s} (\l-x_i)} \ =\ 0,~~~~k=1,\ldots,s-1.
\eeq
It is this kind of boundary condition which leads to the appearance of
hyperelliptic integrals in formulas like \rf{*11}.

\noi
Note that it follows from \rf{*15} that
the planar limit of the spectral density is
\beq{*19}
\rho(\l) = \frac{1}{2\pi} |M(\l)| \sqrt{-\prod_{i=1}^{2s} (\l-x_i)}~~
,~~~~~~~\l\in \sg(\l).
\eeq

\noi
Having obtained the complete solution $G_0(z)$ in the large $N$ limit
one can solve the loop equation  \rf{*12} iteratively as an
expansion in $1/N^2$, and at a given order $h$ it is possible to
construct the multi--point resolvent $G_h(z_1,\ldots,z_n)$ from
$G_h(z)$. The key ingredient in this construction is
the so called {\it loop insertion operator}
\beq{*13a}
\frac{d}{d V(z)} = -\sum_{k} \frac{k}{z^{k+1}} \frac{d}{d g_k}.
\eeq
By definition it follows that\footnote{For $G(z)$ $1/z$ has to be added
on the r.h.s.}
\beq{*20}
G(z_1,\ldots,z_n) = \frac{d}{dV(z_n)}\cdots\frac{d}{dV(z_1)} F,~~~N^2F = 
\log Z,~~~n\geq 2.
\eeq
In particular, we have
\beq{*21}
G(z_1,z_2) = \frac{d}{dV(z_2)} G(z_1),
\eeq
which allows us to construct  $G_0(z_1,z_2)$ from \rf{*15}.
The steps are in principle elementary, but a considerable amount
of algebra is needed in order to prove that $G_0(z_1,z_2)$ is
universal, and of the form \rf{*11} for the example of two 
cuts with an even
potential $V(z)$. The details will be published elsewhere \cite{gernot}.
Let us only mention here that the main complication compared to
the single cut case (i.e. $s=1$) is that the kernel of $\hK - 2G_0(z)$ is
not zero but given by
\beq{*22}
{\rm Ker} ( \hK - 2 G_0(z)) =
{\rm Span}\{ \frac{z^l}{\sqrt{\prod_{i=1}^{2s}(z-x_i)}},\ l=0,\ldots,,s-2\},
\eeq
after taking into account the asymptotic of $G(z)$.

\noi
The $s$-cut solution will be
characterized by $2s$ classes of so called {\it moments}, given by
\beq{*23}
M_i^{(l)} = \frac{1}{(l-1)!} \; \frac{ d^{l-1}}{d\l^{l-1}} M(\l )
\Bigr|_{\l =x_i},~~~i=1,\ldots,2s.
\eeq
Following the derivation for $s=1$, but  with considerable
algebraic complications, one then can show by iteration that
$G_h (z_1,\ldots,z_n)$ can be written in the form
\beq{*23a}
G_h (z_1,\ldots,z_n)  =
\frac{R_h\Big( \{z_j-x_i\}, \{M_i^{(l)}\}, \{x_i\}\Big)}
{\sqrt{\prod_{i=1}^{2s}\prod_{j=1}^n
(z_j-x_i)}},
\eeq
where $R_h$ is a rational function of the poles and the moments
with $1 \leq l \leq 3h-2+n$. However, contrary to the case $s=1$
the coefficients in $R_h$ are no longer rational functions of the $x_i$,
but involve hyperelliptic integrals as well.

\section{Comments}
The universal features of the single cut solution of the random matrix
models generalize to an arbitrary number of cuts. As was shown in
\cite{ackm,leo1,leo2} the universality of the one-cut solution is deeply
connected to the theory of intersection indices on the moduli space of
punctured Riemann surfaces. It is unknown if a similar mathematical
underlying structure can be related to the multi-cut solution.
The fact the the coefficients
in \rf{*23a} can be expressed in terms of hyperelliptic integrals
gives some hope in this direction, since similar hyperelliptic
integrals are encountered for the random matrix model coupled to
$O(n)$ spins and this model is known to be related to integrable
hierarchies.


\begin{thebibliography}{99}
\bibitem{ajm} J. Ambj\o rn, J. Jurkiewicz and Y. Makeenko,
Phys.Lett B251 (1990) 517.
\bibitem{bz} E. Brezin and A. Zee, Nucl.Phys. B402 (1993) 613.
\bibitem{ackm}J. Ambj\o rn, L. Chekhov, C.F. Kristjansen and Yu. Makeenko,
Nucl.Phys. B404 (1993) 127.
\bibitem{superm} J.C. Plefka, Nucl.Phys. B444 (1995) 333; B448 (1995) 355.
\bibitem{akm}J. Ambj\o rn, C.F. Kristjansen and Yu. Makeenko,
Mod.Phys.Lett. A7 (1992) 3187.
\bibitem{david} F. David , Nucl.Phys. B348, (1991) 507.
\bibitem{jurek} J. Jurkiewicz, Phys.Lett. B245 (1990) 178.
\bibitem{gernot} G. Akemann, {\it Higher genus correlators for the hermitian
matrix model with multiple cuts}, NBI-HE-96-30, IPT-UH-08/96, hep-th/9606004.
\bibitem{leo1}L. Chekhov, Geometry and Physics, 12 (1993) 153.
\bibitem{leo2}L. Chekhov, {\it Matrix models and geometry of moduli spaces},
Steklov Mathematical Institute Preprint, hep-th/9509001.
\end{thebibliography}
\end{document}